\documentclass[a4paper,11pt]{article}
\usepackage[mathlines]{lineno}
\usepackage{jcappub}

\usepackage[table,svgnames,dvipsnames]{xcolor}

\usepackage[normalem]{ulem}
\usepackage{aas_macros}
\usepackage{subfigure}
\usepackage{graphicx}
\usepackage{graphics}
\usepackage{dcolumn}
\usepackage{bm}
\usepackage{orcidlink}
\usepackage{cases}
\usepackage{booktabs}
\usepackage{comment}
\usepackage{multirow}
\usepackage{makecell}
\usepackage{siunitx}
\usepackage{tabularx}
\usepackage{xspace}
\usepackage{soul} 
\usepackage{tablefootnote}
\graphicspath{{figs/}}

\usepackage{orcidlink} 
\usepackage{hanging} 
\usepackage{arydshln} 
\usepackage{makecell} 

\newcommand{\tbd}[1]{\textcolor{black}{#1}}
\newcommand{\refedit}[1]{\textcolor{black}{#1}}  
\newcommand{\cwr}[1]{{#1}}
\newcommand{\cwrminor}[1]{{#1}}
\newcommand{\figref}[1]{Fig.~\ref{fig:#1}}
\newcommand{\tabref}[1]{Table~\ref{table:#1}}
\newcommand{\secref}[1]{Section~\ref{sec:#1}}

\def\be{\begin{equation}}
\def\ee{\end{equation}}

\def\ba#1\ea{\begin{align*}#1\end{align*}}

\renewcommand{\emph}[1]{\textit{#1}}
\definecolor{RoyalBlue}{rgb}{0.25,.41,.88}
\definecolor{WildStrawberry}{HTML}{EE2967}
\definecolor{RedWine}{rgb}{0.743,0,0}
\definecolor{bittersweet}{rgb}{1.0, 0.44, 0.37}
\definecolor{burntorange}{rgb}{0.8, 0.33, 0.0}
\definecolor{midnightgreen}{rgb}{0.0, 0.29, 0.33}
\definecolor{otherblue}{rgb}{0.20, 0.73, 0.92}

\usepackage[nameinlink,noabbrev]{cleveref}
\crefname{equation}{Eq.}{Eqs.}
\crefname{section}{Section}{Sections}
\crefname{figure}{Figure}{Figures}
\crefname{table}{Table}{Tables}
\crefname{appendix}{Appendix}{Appendices}
\Crefname{figure}{Figure}{Figures}
\Crefname{equation}{Equation}{Equations}
\Crefname{section}{Section}{Sections}
\Crefname{table}{Table}{Tables}

\newcommand{\mksym}[1]{\ifmmode {\rm #1}\else #1\fi}

\newcommand{\Om}{\Omega_\mathrm{m}}

\newcommand{\Ob}{\Omega_\mathrm{b}}

\newcommand{\ob}{\omega_\mathrm{b}}
\newcommand{\onu}{\omega_\mathrm{\nu}}
\newcommand{\ocb}{\omega_\mathrm{cb}}
\newcommand{\Ocb}{\Omega_\mathrm{cb}}
\newcommand{\ocdm}{\omega_\mathrm{cdm}}

\newcommand{\lcdm}{$\Lambda$CDM\xspace}

\newcommand{\lya}{Ly$\alpha$\xspace}

\newcommand{\rd}{r_\mathrm{d}}

\newcommand{\cs}{c_\mathrm{s}}

\newcommand{\sumnu}{\sum m_\nu}

\newcommand{\Planck}{\emph{Planck}\xspace}

\newcommand{\rs}{r_{\mathrm{s}}}
\newcommand{\keq}{k_{\mathrm{eq}}}
\newcommand{\logA}{\ln{\left(10^{10} 
A_\mathrm{s}\right)}}
\newcommand{\ns}{n_{\mathrm{s}}}
\newcommand{\qbao}{q_{\mathrm{BAO}}}

\newcommand{\hinvmpc}{\,h^{-1}{\rm Mpc}}
\newcommand{\hmpcinv}{\,h\,{\rm Mpc^{-1}}}
\newcommand{\kmsMpc}{\,{\rm km\,s^{-1}\,Mpc^{-1}}}
\newcommand{\eV}{{\,\rm eV}}

\newcommand{\planckact}{\emph{Planck}+ACT\xspace}

\title{A Sound Horizon-Free Measurement of $H_0$ in DESI 2024}

\author[1,2,3]{{E.~A.~Zaborowski}\orcidlink{0000-0002-6779-4277},}
\author[3]{{P.~Taylor},}
\author[1,2,3]{{K.~Honscheid}\orcidlink{0000-0002-6550-2023},}
\author[4]{{A.~Cuceu}\orcidlink{0000-0002-2169-0595},}
\author[5]{{A.~de~Mattia}\orcidlink{0000-0003-0920-2947},}
\author[6,7]{{D.~Huterer}\orcidlink{0000-0001-6558-0112},}
\author[8,9,10]{{A.~Krolewski},}
\author[1,11,3]{{P.~Martini}\orcidlink{0000-0002-4279-4182},}
\author[1,11,3]{{A.~J.~Ross}\orcidlink{0000-0002-7522-9083},}
\author[3]{{C.~To},}
\author[3]{{A.~Torres},}
\author[12]{{S.~Ahlen}\orcidlink{0000-0001-6098-7247},}
\author[13]{{D.~Bianchi}\orcidlink{0000-0001-9712-0006},}
\author[14]{{D.~Brooks},}
\author[15,16]{{E.~Buckley-Geer},}
\author[5]{{E.~Burtin},}
\author[4]{{T.~Claybaugh},}
\author[17]{{S.~Cole}\orcidlink{0000-0002-5954-7903},}
\author[18]{{A.~de la Macorra}\orcidlink{0000-0002-1769-1640},}
\author[19]{{Arjun~Dey}\orcidlink{0000-0002-4928-4003},}
\author[20,21]{{Biprateep~Dey}\orcidlink{0000-0002-5665-7912},}
\author[14]{{P.~Doel},}
\author[4,22]{{S.~Ferraro}\orcidlink{0000-0003-4992-7854},}
\author[14,23]{{A.~Font-Ribera}\orcidlink{0000-0002-3033-7312},}
\author[24,25]{{J.~E.~Forero-Romero}\orcidlink{0000-0002-2890-3725},}
\author[26,27,28]{{E.~Gaztañaga},}
\author[29,26,30]{{H.~Gil-Mar\'in}\orcidlink{0000-0003-0265-6217},}
\author[16]{{G.~Gutierrez},}
\author[4]{{J.~Guy}\orcidlink{0000-0001-9822-6793},}
\author[31]{{C.~Hahn}\orcidlink{0000-0003-1197-0902},}
\author[32]{{C.~Howlett}\orcidlink{0000-0002-1081-9410},}
\author[19]{{S.~Juneau},}
\author[33]{{R.~Kehoe},}
\author[34]{{D.~Kirkby}\orcidlink{0000-0002-8828-5463},}
\author[4]{{T.~Kisner}\orcidlink{0000-0003-3510-7134},}
\author[4]{{A.~Kremin}\orcidlink{0000-0001-6356-7424},}
\author[4]{{M.~Landriau}\orcidlink{0000-0003-1838-8528},}
\author[35]{{L.~Le~Guillou}\orcidlink{0000-0001-7178-8868},}
\author[4]{{M.~E.~Levi}\orcidlink{0000-0003-1887-1018},}
\author[5]{{C.~Magneville},}
\author[19]{{A.~Meisner}\orcidlink{0000-0002-1125-7384},}
\author[36,23]{{R.~Miquel},}
\author[37]{{J.~Moustakas}\orcidlink{0000-0002-2733-4559},}
\author[5,4]{{N.~Palanque-Delabrouille}\orcidlink{0000-0003-3188-784X},}
\author[8,9,10]{{W.~J.~Percival}\orcidlink{0000-0002-0644-5727},}
\author[38]{{F.~Prada}\orcidlink{0000-0001-7145-8674},}
\author[39]{{I.~P\'erez-R\`afols}\orcidlink{0000-0001-6979-0125},}
\author[40]{{G.~Rossi},}
\author[41]{{E.~Sanchez}\orcidlink{0000-0002-9646-8198},}
\author[4]{{D.~Schlegel},}
\author[6,7]{{M.~Schubnell},}
\author[42]{{H.~Seo}\orcidlink{0000-0002-6588-3508},}
\author[19]{{D.~Sprayberry},}
\author[7]{{G.~Tarl\'{e}}\orcidlink{0000-0003-1704-0781},}
\author[19]{{B.~A.~Weaver},}
\author[43,44,45]{{R.~H.~Wechsler}\orcidlink{0000-0003-2229-011X},}

\affiliation[1]{Center for Cosmology and AstroParticle Physics, The Ohio State University, 191 West Woodruff Avenue, Columbus, OH 43210, USA}
\affiliation[2]{Department of Physics, The Ohio State University, 191 West Woodruff Avenue, Columbus, OH 43210, USA}
\affiliation[3]{The Ohio State University, Columbus, 43210 OH, USA}
\affiliation[4]{Lawrence Berkeley National Laboratory, 1 Cyclotron Road, Berkeley, CA 94720, USA}
\affiliation[5]{IRFU, CEA, Universit\'{e} Paris-Saclay, F-91191 Gif-sur-Yvette, France}
\affiliation[6]{Department of Physics, University of Michigan, Ann Arbor, MI 48109, USA}
\affiliation[7]{University of Michigan, Ann Arbor, MI 48109, USA}
\affiliation[8]{Department of Physics and Astronomy, University of Waterloo, 200 University Ave W, Waterloo, ON N2L 3G1, Canada}
\affiliation[9]{Perimeter Institute for Theoretical Physics, 31 Caroline St. North, Waterloo, ON N2L 2Y5, Canada}
\affiliation[10]{Waterloo Centre for Astrophysics, University of Waterloo, 200 University Ave W, Waterloo, ON N2L 3G1, Canada}
\affiliation[11]{Department of Astronomy, The Ohio State University, 4055 McPherson Laboratory, 140 W 18th Avenue, Columbus, OH 43210, USA}
\affiliation[12]{Physics Dept., Boston University, 590 Commonwealth Avenue, Boston, MA 02215, USA}
\affiliation[13]{Dipartimento di Fisica ``Aldo Pontremoli'', Universit\`a degli Studi di Milano, Via Celoria 16, I-20133 Milano, Italy}
\affiliation[14]{Department of Physics \& Astronomy, University College London, Gower Street, London, WC1E 6BT, UK}
\affiliation[15]{Department of Astronomy and Astrophysics, University of Chicago, 5640 South Ellis Avenue, Chicago, IL 60637, USA}
\affiliation[16]{Fermi National Accelerator Laboratory, PO Box 500, Batavia, IL 60510, USA}
\affiliation[17]{Institute for Computational Cosmology, Department of Physics, Durham University, South Road, Durham DH1 3LE, UK}
\affiliation[18]{Instituto de F\'{\i}sica, Universidad Nacional Aut\'{o}noma de M\'{e}xico,  Circuito de la Investigaci\'{o}n Cient\'{\i}fica, Ciudad Universitaria, Cd. de M\'{e}xico  C.~P.~04510,  M\'{e}xico}
\affiliation[19]{NSF NOIRLab, 950 N. Cherry Ave., Tucson, AZ 85719, USA}
\affiliation[20]{Department of Astronomy \& Astrophysics, University of Toronto, Toronto, ON M5S 3H4, Canada}
\affiliation[21]{Department of Physics \& Astronomy and Pittsburgh Particle Physics, Astrophysics, and Cosmology Center (PITT PACC), University of Pittsburgh, 3941 O'Hara Street, Pittsburgh, PA 15260, USA}
\affiliation[22]{University of California, Berkeley, 110 Sproul Hall \#5800 Berkeley, CA 94720, USA}
\affiliation[23]{Institut de F\'{i}sica d’Altes Energies (IFAE), The Barcelona Institute of Science and Technology, Campus UAB, 08193 Bellaterra Barcelona, Spain}
\affiliation[24]{Departamento de F\'isica, Universidad de los Andes, Cra. 1 No. 18A-10, Edificio Ip, CP 111711, Bogot\'a, Colombia}
\affiliation[25]{Observatorio Astron\'omico, Universidad de los Andes, Cra. 1 No. 18A-10, Edificio H, CP 111711 Bogot\'a, Colombia}
\affiliation[26]{Institut d'Estudis Espacials de Catalunya (IEEC), 08034 Barcelona, Spain}
\affiliation[27]{Institute of Cosmology and Gravitation, University of Portsmouth, Dennis Sciama Building, Portsmouth, PO1 3FX, UK}
\affiliation[28]{Institute of Space Sciences, ICE-CSIC, Campus UAB, Carrer de Can Magrans s/n, 08913 Bellaterra, Barcelona, Spain}
\affiliation[29]{Departament de F\'{\i}sica Qu\`{a}ntica i Astrof\'{\i}sica, Universitat de Barcelona, Mart\'{\i} i Franqu\`{e}s 1, E08028 Barcelona, Spain}
\affiliation[30]{Institut de Ci\`encies del Cosmos (ICCUB), Universitat de Barcelona (UB), c. Mart\'i i Franqu\`es, 1, 08028 Barcelona, Spain.}
\affiliation[31]{Department of Astrophysical Sciences, Princeton University, Princeton NJ 08544, USA}
\affiliation[32]{School of Mathematics and Physics, University of Queensland, 4072, Australia}
\affiliation[33]{Department of Physics, Southern Methodist University, 3215 Daniel Avenue, Dallas, TX 75275, USA}
\affiliation[34]{Department of Physics and Astronomy, University of California, Irvine, 92697, USA}
\affiliation[35]{Sorbonne Universit\'{e}, CNRS/IN2P3, Laboratoire de Physique Nucl\'{e}aire et de Hautes Energies (LPNHE), FR-75005 Paris, France}
\affiliation[36]{Instituci\'{o} Catalana de Recerca i Estudis Avan\c{c}ats, Passeig de Llu\'{\i}s Companys, 23, 08010 Barcelona, Spain}
\affiliation[37]{Department of Physics and Astronomy, Siena College, 515 Loudon Road, Loudonville, NY 12211, USA}
\affiliation[38]{Instituto de Astrof\'{i}sica de Andaluc\'{i}a (CSIC), Glorieta de la Astronom\'{i}a, s/n, E-18008 Granada, Spain}
\affiliation[39]{Departament de F\'isica, EEBE, Universitat Polit\`ecnica de Catalunya, c/Eduard Maristany 10, 08930 Barcelona, Spain}
\affiliation[40]{Department of Physics and Astronomy, Sejong University, Seoul, 143-747, Korea}
\affiliation[41]{CIEMAT, Avenida Complutense 40, E-28040 Madrid, Spain}
\affiliation[42]{Department of Physics \& Astronomy, Ohio University, Athens, OH 45701, USA}
\affiliation[43]{Kavli Institute for Particle Astrophysics and Cosmology, Stanford University, Menlo Park, CA 94305, USA}
\affiliation[44]{Physics Department, Stanford University, Stanford, CA 93405, USA}
\affiliation[45]{SLAC National Accelerator Laboratory, Menlo Park, CA 94305, USA}

\emailAdd{zaborowski.11@osu.edu}

\date{\today}

\abstract{
The physical size of the sound horizon at recombination is a powerful source of information for early-time measurements of the Hubble constant $H_0$, and many proposed solutions to the Hubble tension therefore involve modifications to this scale.
In light of this, there has been growing interest in measuring $H_0$ independently of the sound horizon. 
We present the first such measurement to use data from the Dark Energy Spectroscopic Instrument (DESI), jointly analyzing the full-shape galaxy power spectra of DESI luminous red galaxies, emission line galaxies, quasars, \cwr{and the bright galaxy sample,} in a total of \cwr{six} redshift bins.
Information from the sound horizon scale is removed from our constraints via a rescaling procedure at the power spectrum level, \cwrminor{with our sound horizon-marginalized measurement being driven instead primarily by the matter-radiation equality scale}.
This measurement is then combined with additional sound horizon-free information from \planckact CMB lensing, uncalibrated type Ia supernovae, and the DESI Lyman-$\alpha$ forest.
We agnostically combine with the DESY5, Pantheon+, and Union3 supernova datasets, with our tightest respective constraints being $H_0=\cwr{66.7^{+1.7}_{-1.9}},~\cwr{67.9^{+1.9}_{-2.1}},$ and $\cwr{67.8^{+2.0}_{-2.2}} \,\kmsMpc$.
This corresponds to a sub-3\% sound horizon-free constraint of the Hubble constant, and is the most precise measurement of its kind to date.
Even without including information from the sound horizon, our measurement is still in \cwr{2.2-3.0}$\sigma$ tension with SH0ES.
Additionally, the consistency between our result and other measurements that \textit{do} rely on the sound horizon scale provides no evidence for new early-Universe physics (e.g. early dark energy).
Future DESI data releases will allow unprecedented measurements of $H_0$ and place strong constraints on models that use beyond-\lcdm physics to ameliorate the Hubble tension.
}

\begin{document}
\maketitle
\flushbottom

\section{Introduction}
\label{sec:intro}

The Hubble tension (see \cite{Abdalla:2022} for a review) is one of the most pressing issues in cosmology.
The Hubble constant $H_0$ parameterizes the expansion rate of the Universe, and in particular there is a $\sim$5$\sigma$ discrepancy between the local distance ladder measurement of $H_0$ by the SH0ES collaboration \cite{Riess:2022}, and the measurement reported by the \Planck collaboration \cite{Planck:2020} using the cosmic microwave background (CMB).
This difference could arise either because of unrealized systematic errors in one or both measurements, or due to new physics beyond the standard $\Lambda$ Cold Dark Matter (\lcdm) model of cosmology (see \cite{DiValentino:2021} for an extensive review of possible solutions to the Hubble tension, \refedit{and e.g. \cite{Wojtak:2022,Wojtak:2024,Christa:2024} for more recent investigations of some systematics}).
Many high-redshift/early-time $H_0$ constraints, such as those from the CMB (e.g. \cite{Planck:2020,Madhavacheril:2024}) or from baryon acoustic oscillations (BAO; e.g. \cite{Alam:2017,Cuceu:2019,Abbott:2022,DESI2024.III.KP4,DESI2024.IV.KP6,DESI2024.VI.KP7A}), derive a significant amount of information from the physics of recombination and the sound horizon distance at the time of recombination.\footnote{We note that there is a subtle distinction between the sound horizon distance at photon decoupling ($\rs\,$, which sources features in the CMB), and the sound horizon distance at baryon decoupling ($\rd\,$, which sources baryonic features in the matter power spectrum) \cite{Hu:1996,Eisenstein:1998}. Where the distinction is not important we refer simply to the ``sound horizon.''}
\cwrminor{The sound horizon is a standard ruler that, once calibrated, can be used to infer $H_0$ from the relevant features in these datasets (see \secref{theory}).}
Because of this, many solutions to the Hubble tension that rely on modifications to physics to decrease the physical size of the sound horizon have been proposed \cite{Knox:2020}.
This includes, among others: early dark energy (EDE) \cite{Doran:2006,Bielefeld:2013,Karwal:2016}, additional light degrees of freedom \cite{Eisenstein:2004,Hou:2013} or new interactions between them \cite{Cyr-Racine:2014,Lancaster:2017,Kreisch:2020}, and reduction of the sound speed in the photon-baryon plasma prior to decoupling \cite{Knox:2020}.

For this reason, there has been growing interest in measuring $H_0$ independently of the sound horizon scale \cite{Baxter:2020,Philcox:2021,Brieden:2023,Smith:2023}.
In \cite{Baxter:2020} it was proposed to measure $H_0$ using the gravitational lensing of the CMB, which is not sensitive to the sound horizon scale, but rather to the wavenumber $\keq$ corresponding to the comoving horizon size at the epoch of matter-radiation equality (see \secref{theory}).
The precision of this measurement can then be improved by combining with additional information on the matter density $\Om$, which they sourced from uncalibrated type Ia supernovae (SNe Ia). 
A recent application of this method by \cite{Farren:2024} using CMB lensing data from the Atacama Cosmology Telescope (ACT; \cite{Das:2011,Sherwin:2017}) and \Planck \cite{Planck:2014, Planck:2016,Planck:2020lens, Carron:2022}, combined with galaxies from the unWISE catalog \cite{Schlafly:2019} and uncalibrated supernovae from the Pantheon+ dataset \cite{Brout:2022}, resulted in a precise sound horizon-free measurement of $H_0 = 64.3^{+2.1}_{-2.4} \kmsMpc$.
Others have measured $H_0$ without the sound horizon through a direct measurement of $\keq$, via the turnover scale of the matter power spectrum \cite{Bahr-Kalus:2023}.
In \cite{Prada:2011}, a similar measurement of the horizon size at matter-radiation equality was proposed via the zero-crossing scale of the matter correlation function.
Recently, \cite{Krolewski:2024} demonstrated that $H_0$ can be derived from measurements of cosmological energy densities without including information from the sound horizon scale.

Galaxy surveys provide one of the most promising avenues to measure $H_0$ without the sound horizon.
As remarked in \cite{Philcox:2021}, a 3D galaxy survey can typically probe a much larger number of modes than CMB lensing, and ongoing and upcoming surveys (e.g. the Dark Energy Spectroscopic Instrument (DESI; \cite{DESI2016a.Science, DESI2016b.Instr,DESI2023a.KP1.SV}), Euclid \cite{Euclid:2024}, the Nancy Grace Roman Space Telescope \cite{Wang:2022}, and SPHEREx \cite{Dore:2015}) are pushing to ever-larger volumes.
In \cite{Philcox:2021}, using galaxy power spectrum measurements from the completed Baryon Oscillation Spectroscopic Survey (BOSS DR12; \cite{Dawson:2012,Beutler:2016}),  the authors obtained a measurement of the Hubble constant without the sound horizon by using an uninformative prior on the physical baryon density $\ob \equiv \Ob h^2$.
This method has the benefit of simplicity and leaves the physical size of the sound horizon uncalibrated, but as we explain in \secref{theory}, $\ob$ also contributes to multiple physical effects that constrain $H_0$, and removing this information is not ideal.
In \cite{Farren:2022}, a technique was demonstrated to rescale and marginalize over the sound horizon scale during inference, thus allowing the use of an informative $\ob$ prior while removing information from the sound horizon scale.
Building on this technique, \cite{Philcox:2022} obtained a 3.6\% measurement of the Hubble constant by combining BOSS power spectra with \Planck CMB lensing and SNe Ia from the Pantheon+ dataset.

In this work, we perform a similar (but blind) measurement to that of \cite{Philcox:2022}, using data from the first public data realease of the Dark Energy Spectroscopic Instrument (DESI 2024; \cite{DESI2024.I.DR1}).
\cwr{We note that we performed our analysis ``blind'' in the sense that we only tested our pipeline using mocks before finally running on real data.}
We combine our result with CMB lensing data from ACT \cite{Das:2011,Sherwin:2017} and \Planck \cite{Planck:2014, Planck:2016,Planck:2020lens, Carron:2022}.
We further increase the precision of our measurement by including additional $\Om$ information from two sources: the Alcock-Paczy\'{n}ski (AP) effect \cite{Alcock:1979} observed in the DESI 3D Lyman-$\alpha$ (\lya) forest \cite{Cuceu:2024}, and uncalibrated type Ia supernovae.
This work represents the most precise measurement of its kind to date, achieving a sub-3\% sound horizon-free measurement of $H_0$.
The remainder of this paper is laid out as follows.
In \secref{theory} we discuss how each dataset used in this work helps to constrain the Hubble constant without including information from the sound horizon scale, and then in \secref{data} we describe each specific dataset in detail.
In \secref{methodology} we describe the steps and components of our analysis pipeline.
We present the results of our analysis in \secref{results}, followed by a discussion in \secref{discussion}.
Finally, we conclude in \secref{conclusions}.

\section{Theory and Methodology}
\label{sec:theory}

The galaxy power spectrum contains multiple sources of information that contribute to measuring $H_0$.
On one hand there is geometrical information; one form that this takes is in so-called ``standard rulers,'' or features of the power spectrum that correspond to theoretically known distance scales.
This includes the sound horizon scale at baryon-photon decoupling $\rd$, which determines the structure of the BAO wiggles, as well as the baryonic Jeans suppression scale \cite{Lesgourgues:2006}.
Another standard ruler is the matter-radiation equality scale $\keq$, which manifests as the turnover scale of the power spectrum, and also modulates its shape\footnote{As discussed in \cite{Philcox:2021}, $\keq$ appears non-trivially in the shape of the logarithmic enhancement on scales $k > \keq$.} \refedit{and impacts its amplitude on smaller scales \cite{Smith:2023}}.
The observed scale (in angles/redshifts) of a given standard ruler is then a function of both its physical size as well as the distance to the tracer it is measured from, both of which are predicted by a cosmological model.
This leads to a measurement of the relevant cosmological parameters (or combinations thereof), \refedit{where the Hubble constant in particular is inversely proportional to the implied physical distances}.
Further geometrical information is gained by jointly analyzing tracers at multiple redshifts.
Because $\keq \propto \Om h$ (when measured in $\hmpcinv$ units; see Equation \ref{eq:keq}), constraints based on the matter-radiation equality scale $\keq$ display a characteristic degeneracy in the $\Om - h$ plane.
\cwrminor{However, because we observe power spectrum features in angles ($\theta_{\mathrm{eq}} \sim \ell_{\mathrm{eq}}^{-1} \sim \left(D_{\mathrm{A}}\left(z\right) \keq\right)^{-1}$) and redshifts ($\Delta z_{\mathrm{eq}} \sim H\left(z\right) / \keq$) rather than physical distances, this degeneracy direction varies with redshift according to the assumed fiducial cosmological model.}
The resulting constraints, while \cwrminor{largely} degenerate for any individual tracer, then intersect at different angles, resulting in a more precise joint constraint \refedit{(assuming they are consistent and can be combined)}.
A similar phenomenon arises in BAO analyses (see for example Figure 2 of \cite{DESI2024.VI.KP7A}).

Somewhat related to geometrical information is the Alcock-Paczy\'{n}ski (AP) effect \cite{Alcock:1979}.
\cwrminor{When a fiducial cosmology is used to transform galaxy positions from redshift space (R.A., Dec., z) to Cartesian space (x, y, z), any difference between the assumed cosmology and true cosmology will cause clustering features to appear stretched along/across the line of sight.
This fictitious distortion is known as the AP effect, and
can be related to the true matter density $\Om$ (e.g. \cite{Cuceu:2024}).}
Importantly, the AP distortion is independent of absolute distance scales, and thus we can use it to help to break the $\Om - H_0$ degeneracy alluded to earlier.
\cwrminor{Measuring the AP effect requires modelling the full anisotropic contribution to the clustering signal, known as redshift space distortions (RSD), where the latter are also induced by peculiar velocities related to the growth of structure.
Because of its dependence on the peculiar velocity field, the RSD measurement can be used to obtain information about the amplitude of the power spectrum, which also has some degeneracy with the Hubble constant (see \secref{results}).}
Beyond geometric sources, the AP effect, and RSD, note that there is still additional information contained in the shape of the power spectrum. 
For example, \cite{Krolewski:2024} demonstrated a method to derive $H_0$ using energy density measurements, making use of (among other measurements) the amplitude of the BAO signal from galaxy clustering while marginalizing over the physical sound horizon scale.

Because we want to capture as much information as possible, we choose to analyze the entire shape of power spectrum, rather than any single feature.
To do this we use a \texttt{Full Modeling} approach (adopting the terminology of \cite{DESI2024.VII.KP7B}), in which we directly generate model power spectrum multipoles from cosmological parameters and fit them to the observed data.
\cwrminor{
This is in contrast to compression-based approaches, in which compressed parameters are fit to the data in a more model-independent way.
A ubiquitous example of this is the $\alpha_\parallel$, $\alpha_\perp$ (or $\alpha_{\mathrm{iso}}$, $\alpha_{\mathrm{AP}}$) parameterization used in BAO analyses (e.g. \cite{DESI2024.III.KP4}).
Recently, there have been extensions of this approach to full-shape analyses, e.g. \texttt{ShapeFit} \cite{Brieden:2021}, which uses a template power spectrum to fit for individual effects (RSD, Alcock-Paczy\'{n}ski, etc.) with respect to the fiducial cosmology (see e.g. \cite{Brieden:2023} for a precise sound horizon-independent measurement of $H_0$ using the ShapeFit approach).
}
While the Full Modeling approach is more computationally expensive, it captures all of the available information without compression (see e.g. \cite{KP5s2-Maus,KP5s3-Noriega,KP5s4-Lai} for comparisons of Full Modeling and compression/template-based techniques).

In \lcdm the sound horizon at the redshift of baryon-photon decoupling $z_\mathrm{d}$ (also known as the baryon drag epoch) is given by:
\begin{align}
    \nonumber
    \rd &\equiv \int_{z_d}^{\infty} \frac{\cs(z)}{H(z)} \,dz \\
    &\approx \frac{55.154 h \exp\left[-72.3\left(\onu + 0.0006\right)^2\right]}{\ocb^{0.25351} \ob^{0.12807}} \, \left[\hinvmpc\right]
	\label{eq:rs}
\end{align}
\cite{Aubourg:2015}, where $\cs$ is the speed of sound in the photon-baryon plasma prior to decoupling, and $\ocb \equiv \ocdm + \ob$.
On the other hand, the matter-radiation equality scale can be calculated as:
\begin{align}
    \keq = \left(2 \Ocb H_0^2 z_{\mathrm{eq}}\right)^{1/2} \approx 7.46 \times 10^{-2} \Ocb h \Theta_{2.7}^{-2} \,\left[\hmpcinv\right]
	\label{eq:keq}
\end{align}
\cite{Eisenstein:1998}, where $\Theta_{2.7} \equiv T_{\mathrm{CMB}} / (2.7\,\mathrm{K})$.
Naturally, the full shape of the power spectrum contains information from both scales, and we would like to remove the sound horizon information from our measurement.
In \cite{Philcox:2021}, it was shown that this can be accomplished simply by performing a Full Modeling analysis with an uninformative prior on $\ob$.
In this way, the physical size of the sound horizon remains uncalibrated,\footnote{The authors noted that in principle, $\rd$ can be self-calibrated from the combination of the BAO wiggles and the baryonic Jeans suppression scale, but this was not a strong enough prior at the precision of BOSS. In \cite{Farren:2022} the authors further demonstrated that this effect is not expected to be significant even for a Euclid-like survey.} and thus cannot be used as a standard ruler in the measurement of $H_0$.
However, this method has the disadvantage of weakening other sources of $H_0$ information that also depend on the baryon density, e.g. the matter-radiation equality scale, or the critical density of the Universe (in the sense of \cite{Krolewski:2024}).
An improved method, demonstrated in \cite{Farren:2022} and subsequently used in \cite{Philcox:2022}, heuristically rescales and marginalizes over $\rd$ while still allowing the use of an informative prior on $\ob$.
We adopt this approach, which we describe below.

\begin{figure}
\begin{center}
	\includegraphics[width=0.8\columnwidth]{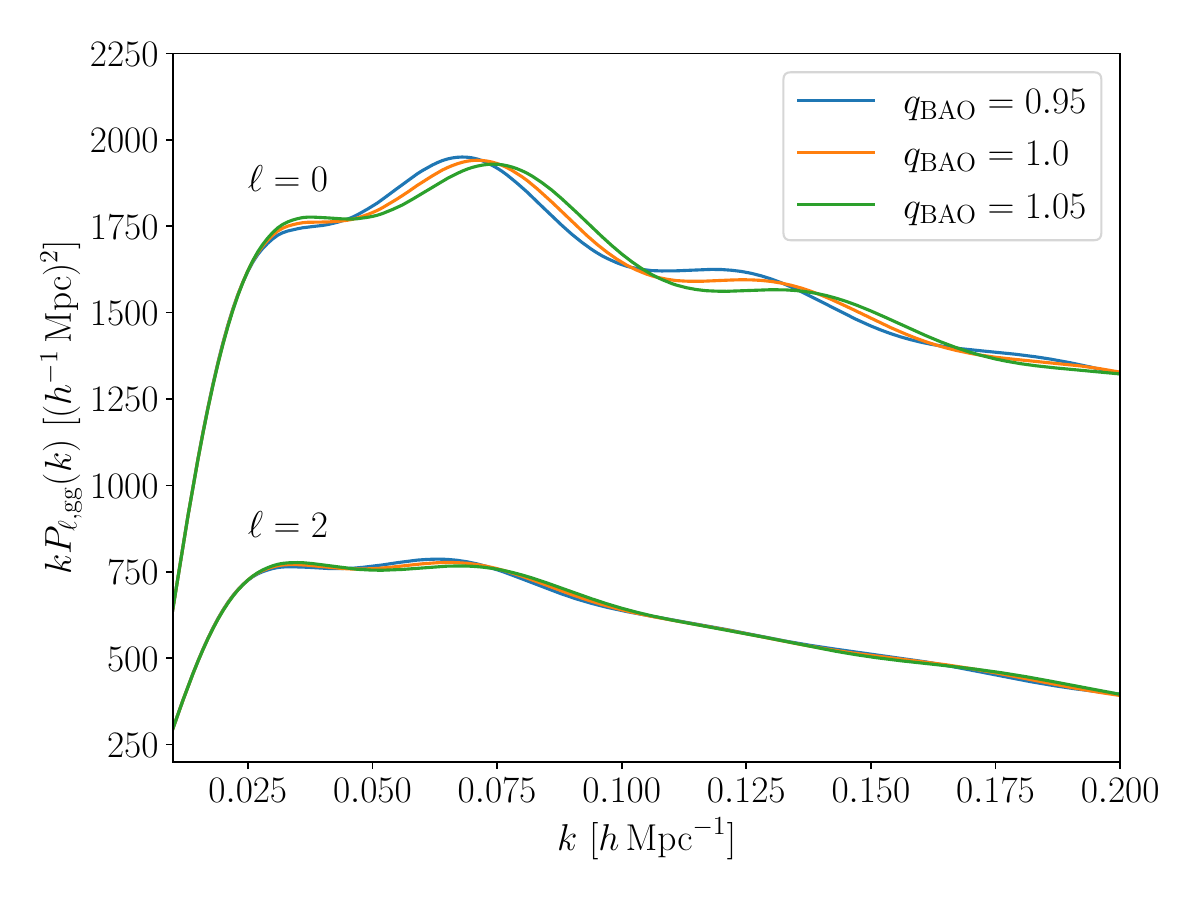}
    \caption{A visual demonstration of the heuristic sound horizon rescaling technique used in this work.
    Adopting the methodology developed in \cite{Farren:2022}, we rescale the power spectrum wiggles by a free parameter $\qbao$. Shown above are the monopole ($\ell=0$) and quadrupole ($\ell=2$) of the theoretical galaxy power spectrum of DESI LRGs in the redshift range $0.4 < z < 0.6$, for three values of $\qbao$.
    $\qbao=1.0$ corresponds to the \lcdm prediction (no rescaling of the sound horizon), whereas $\qbao=1.05$ and $\qbao=0.95$ have the effect of moving the wiggles forward and backward, respectively.
    }
    \label{fig:rescaling}
\end{center}
\end{figure}

We first split the linear power spectrum into a smooth and wiggly component.
To accomplish this, we use a simple version\footnote{https://github.com/cosmodesi/cosmoprimo/blob/main/cosmoprimo/bao\_filter.py\#L472} of the method described in Appendix D of \cite{Brieden:2022} to obtain the smooth part of the power spectrum, where the wiggly part can then be obtained as the difference between the original power spectrum and the smooth part.
\cwrminor{
We refer the reader to the given reference for details of the method; here we give only a brief summary.
The broadband shape of the power spectrum is first divided out using the smooth Eisenstein-Hu formula \cite{Eisenstein:1998}, resulting in an approximation of the wiggles.
Then, the gradient of the latter is calculated numerically and extremized.
The maxima and minima thus obtained correspond to the nodes of the wiggles; importantly, the smooth part must also pass through these nodes.
Two smooth functions are fit using quadratic splines: one each crossing the nodes that correspond to the gradient maxima or minima, and these curves are finally averaged to give the smooth part of the power spectrum.
}

Once the smooth and wiggly components have been separated using the above technique, we rescale only the wiggly part by a new parameter $\qbao$ (called $\alpha_{\rs}$ in \cite{Farren:2022}).
We then recombine the two components, such that the transformed linear power spectrum can be described as:
\begin{align}
    P_{\mathrm{lin}}\left(k,\qbao\right) = P_{\mathrm{lin}}^{\mathrm{smooth}}\left(k\right) + P_{\mathrm{lin}}^{\mathrm{wiggly}}\left(\qbao \cdot k\right)
\end{align}
We can then use this modified linear power spectrum as an input to our full non-linear galaxy clustering model, described in \secref{theory_code}.
Roughly speaking, the effect of this heuristic transformation is to rescale $\rd$ by a factor $\qbao^{-1}$, phenomenologically capturing any modifications to the physics that determines the scale of this feature.
In \figref{rescaling} we show visually the rescaling of the galaxy power spectrum monopole and quadrupole for three values of $\qbao$.
Note that $\qbao=1.0$ corresponds to \lcdm (no rescaling), while larger (smaller) values of $\qbao$ shift the BAO wiggles forward (backward).
By allowing $\qbao$ to vary freely, the information from the sound horizon scale is marginalized over during inference.
We note that several tests were performed in \cite{Farren:2022} and \cite{Philcox:2022} to demonstrate that the constraints obtained from this procedure are indeed independent of sound horizon information, at least at the precision of DESI and Euclid.

As mentioned earlier, the CMB lensing power spectrum is also sensitive to the matter-radiation equality scale.
As discussed in \cite{Baxter:2020}, the projection integral relating the lensing power spectrum to the intervening matter power spectrum has the effect of washing out sharp features such as the BAO wiggles.
Because of this, CMB lensing is not sensitive to the sound horizon scale,\footnote{In \cite{Baxter:2020}, the authors additionally verify that no significant information is sourced from the baryonic Jeans suppression scale.} but rather to the broadband shape of the matter power spectrum, which is modulated by the matter-radiation equality scale as noted above.
Thus CMB lensing constrains $H_0$ similarly to above, albeit with a distinct degeneracy direction in the $\Om - H_0$ plane, helping to improve the joint constraint.

Clearly, the $\Om - H_0$ degeneracy can be broken further by including additional information on $\Om$.
In this work we consider two additional sources of $\Om$ information.
The first of these is type Ia supernovae (SNe Ia).
SNe Ia are standardizable candles that trace the expansion history of the Universe.
When the local distance ladder \cite{Riess:2022,Freedman:2024} is used to calibrate the absolute magnitudes of SNe Ia, their observed brightnesses imply physical distances, and they can be used to measure $H_0$.
When left uncalibrated they no longer provide absolute distance information, but still probe the expansion history.
In a flat universe and for a fixed dark energy equation of state, this primarily provides information on $\Om$.
The second source of $\Om$ information we include in our measurement is the AP effect observed in the full-shape 3D Lyman-$\alpha$ forest power spectrum \cite{Cuceu:2021,Cuceu:2023,Cuceu:2024}.
As mentioned above, information from the AP effect is also included in galaxy clustering, however the effect is stronger for the \lya forest as it is more sensitive to small-scale structures \cite{Cuceu:2021}.
The \lya forest \cite{Lynds:1971,Rauch:1998} can be seen in the spectra of distant quasars, and traces the density of the intervening neutral hydrogen.
As light travels toward us from a quasar and becomes redshifted by the expansion of space, it can be absorbed by neutral hydrogen (in this case as \lya photons).
Thus we observe a ``forest'' of absorption lines in the quasar spectrum, blueward of the quasar's rest-frame \lya wavelength.
By considering the emitted wavelength of the quasar light and the strength of its absorption, the structure of the intervening neutral hydrogen can be inferred.
The AP distortion observed in the \lya forest power spectrum then provides information on $\Om$, as 
discussed previously.

\section{Data}
\label{sec:data}

Here we describe each specific dataset used in this work.

\subsection{Galaxies and the Lyman-$\alpha$ Forest}

The Dark Energy Spectroscopic Instrument (DESI) is a multi-object spectrographic survey performed using $\sim$5,000 robotically positioned fibers and operating on the Mayall 4-meter telescope at Kitt Peak National Observatory \cite{DESI2022.KP1.Instr,DESI2016b.Instr,FocalPlane.Silber.2023,Corrector.Miller.2023}.
DESI is currently conducting a five-year survey that will cover roughly 14,000 $\deg^2$ of the sky and obtain the spectra of more than 40 million galaxies and quasars \cite{DESI2016a.Science}.
The main science goal of the survey is to probe the nature of dark energy via the most precise measurement of the expansion history of the Universe to date \cite{Snowmass2013.Levi}.
The spectroscopic data reduction pipeline used for DESI is described in \cite{Spectro.Pipeline.Guy.2023}, while the survey operations and observation plan are detailed in \cite{SurveyOps.Schlafly.2023}.
DESI target selection is based on imaging data from the public DESI Legacy Imaging Surveys \cite{LS.Overview.Dey.2019}.
The DESI scientific program was validated in \cite{DESI2023a.KP1.SV}, accompanied by an early data release \cite{DESI2023b.KP1.EDR}.

The first-year data release of the DESI survey (DESI 2024; \cite{DESI2024.I.DR1}) contains the positions and spectra of over 5.7 million unique objects, which are classified among several tracers in multiple redshift bins.
Accompanying the DESI 2024 data release are several science Key Papers: two-point clustering measurements and validation \cite{DESI2024.II.KP3}, measurements of the BAO feature in galaxies and quasars \cite{DESI2024.III.KP4} and the Lyman-$\alpha$ (\lya) forest \cite{DESI2024.IV.KP6}, and the full-shape study of galaxies and quasars \cite{DESI2024.V.KP5}.
Cosmological interpretations were given for the BAO measurements \cite{DESI2024.VI.KP7A} and the full-shape analysis \cite{DESI2024.VII.KP7B}.
In this work we jointly analyze the full-shape clustering of a total of \cwr{six} unique tracer/redshift combinations in DESI 2024: \cwr{the low-redshift bright galaxy sample (BGS) at redshift range $0.1 < z < 0.4$}, luminous red galaxies (LRGs) in three redshift bins ($0.4 < z < 0.6$, $0.6 < z < 0.8$, and $0.8 < z < 1.1$), emission line galaxies (ELGs) in the redshift range $1.1 < z < 1.6$, and quasars (quasi-stellar objects or QSOs) in the redshift range $0.8 < z < 2.1$, together comprising a total of 4.7 million galaxies.
Additionally, we make use of measurements of the \lya forest, as observed in the spectra of distant DESI quasars.
In particular, we use the measurement by \cite{Cuceu:2024} of the Alcock-Paczy\'{n}ski effect observed in the 3D power spectrum of the DESI \lya forest.

\subsection{CMB Lensing}
We use a combination of CMB lensing data from the \Planck \cite{Planck:2020} and ACT \cite{Madhavacheril:2024,Qu:2024} surveys. 
Specifically, we use the \Planck \textsc{NPIPE} PR4 lensing reconstruction \cite{Carron:2022}, which covers $\sim$27,600 $\deg^2$ of the sky, and the ACT DR6 lensing reconstruction \cite{Madhavacheril:2024,Qu:2024}, covering 9,400 $\deg^2$.
We make use of the publicly available\footnote{https://github.com/ACTCollaboration/act\_dr6\_lenslike\,. We enforced the following settings for our analysis: \texttt{variant = actplanck\_baseline} and \texttt{lens\_only = true}\,.} \planckact CMB-marginalized lensing-only likelihood.

\subsection{Type Ia Supernovae}
As in \cite{DESI2024.VI.KP7A} and \cite{DESI2024.VII.KP7B}, we combine our results with three separate uncalibrated supernova datasets.
Because uncalibrated SNe Ia primarily constrain $\Om$ (as was noted in \secref{theory}), we choose to extract $\Om$ priors from each dataset rather than working with the full posteriors, which would provide little additional information.
We perform the combination with each supernova dataset individually, without favoring any one over the other.
These datasets are the Dark Energy Survey Year 5 supernova analysis (DESY5; \cite{DES:2024}), Pantheon+ \cite{Scolnic:2022}, and Union3 \cite{Rubin:2023}.
We very briefly describe each dataset, referring the reader to the relevant references for further information.
DESY5 represents the largest and most homogeneous single sample of high-redshift supernovae to date; it combines novel observations of 1,635 photometrically classified SNe Ia (spanning $0.1 < z < 1.3$) with existing observations of 194 low-redshift ($0.025 < z < 0.1$) SNe Ia, the latter of which are also shared by Pantheon+.
The Pantheon+ dataset itself is a compilation of data from many surveys; it contains four separate high-redshift samples ($z > 1.0$), three mid-redshift samples ($0.1 < z < 1.0$), and 11 different low-redshift samples ($z < 0.1$), for a total of 1,550 spectroscopically confirmed SNe Ia.
Union3 consists of 2,087 SNe Ia, sharing 1,363 with Pantheon+, and additionally makes use of a Bayesian Hierarchical Modeling approach in its likelihood analysis.

\section{Likelihood and Analysis}
\label{sec:methodology}

Here we describe the settings and techniques used in our analysis. The entire analysis pipeline is built into the code \texttt{desilike}.\footnote{https://desilike.readthedocs.io}
Throughout this work we assume a cosmology with $\tau_{\mathrm{reio}}=0.0544$, $N_{\mathrm{eff}}=3.046$, and three degenerate species of massive neutrinos \refedit{as in \cite{Planck:2020}}.

\subsection{Data Vector}
\label{sec:datavec}

We filter the galaxy power spectrum multipoles measured in DESI 2024 for each tracer to wavenumber $k$ in the range $\left[0.02, 0.20\right] \hmpcinv$ for $\ell=0,2$, in steps $\mathrm{d}k=0.005 \hmpcinv$.
This is consistent with the analyses of \cite{DESI2024.V.KP5,DESI2024.VII.KP7B}, however we note that the $k$-range of our data vector does not include the expected turnover scale of the power spectrum, $\keq \sim 0.015 \hmpcinv$.
As remarked in \secref{theory}, the matter-radiation equality scale not only sets the turnover scale of the power spectrum, but also appears non-trivially in its broadband shape.
In fact, \cite{Smith:2023} showed that for data from BOSS, the scales $k > \keq$ are dominant over the position of the turnover scale in constraining $H_0$ without the sound horizon.
We expect this also to be the case for the first DESI data release, but this may change in the future as the peak of the power spectrum  becomes better resolved.

\cwrminor{
All additional processing of the data, including truncation of small angular scales \cite{Pinon:2024}, ``rotation'' of the data vector to compactify the window matrix in $k$-space \cite{Pinon:2024}, and corrections for the radial integral constraint (RIC) \cite{DeMattia:2019} and angular integral constraint (AIC) \cite{DESI2024.II.KP3}, directly follows the methodology described in \cite{DESI2024.II.KP3}.
The latter two effects (the RIC and AIC) are particularly important to account for in this analysis, as they lead to the damping of power on large radial and angular scales, respectively.
In summary, the RIC arises because the redshift distribution of randoms exactly matches that of the data (by design), while the AIC is caused by the use of imaging systematic weights that are obtained by regressing the galaxy density on several maps that encode various imaging properties.
Both effects are individually corrected by comparing the mean power spectra of mocks that do/do not contain the particular effect, and fitting a polynomial to capture the difference in power.
}

\subsection{Covariance Matrix}
\label{sec:covariance}

We use a covariance matrix that is a sum of both a statistical and systematic component.
\cwrminor{
The statistical part of the covariance matrix is described in detail in \cite{KP4s6-Forero-Sanchez}, whereas the systematic contributions are given in \cite{DESI2024.V.KP5}, with the detailed methodology of their combination shown in Appendix D of the latter.
}
The statistical part of covariance matrix was calculated using 1,000 EZmocks (effective Zel'dovich approximation mocks, \cite{Chuang:2015}), which are faster to create than full $N$-body simulations, while still accurately capturing the DESI 2024 clustering statistics. 
\cwrminor{
As discussed in \cite{DESI2024.V.KP5}, any systematic effect that leads to a parameter-level bias of at least $0.2\sigma$ (in units of the DESI 2024 error) is included in the systematic component of the covariance matrix.
These systematic contributions include imaging effects, fiber assignment, and halo occupation distribution (HOD)-dependent systematics.
Once added in quadrature, these effects contribute a systematic uncertainty of $\sim$0.46$\sigma$ in units of the DESI 2024 error, which is much smaller than the sound horizon-marginalized error level of this work.
}

\subsection{Power Spectrum Model}
\label{sec:theory_code}
To calculate the theoretical galaxy power spectrum, we use the publicly available code \texttt{velocileptors}\footnote{https://github.com/sfschen/velocileptors} \cite{Chen:2020,Chen:2021}, which is based on the effective field theory of large-scale structure (EFT of LSS).
We use the Lagrangian perturbation theory (LPT) implementation up to third order, fixing the third-order bias parameter $b_3$ to zero similarly to \cite{DESI2024.V.KP5,DESI2024.VII.KP7B}.
We adopt a physically motivated reparameterization of the EFT biases, counterterms, and stochastic terms, which is outlined in \cite{DESI2024.V.KP5}.
By using this parameterization, physically interpretable priors can be enforced on these parameters, as discussed in \secref{priors}.

\subsection{Priors}
\label{sec:priors}

Our prior choices largely follow \cite{DESI2024.V.KP5,DESI2024.VII.KP7B}, and we summarize them below while noting any differences.
The baseline Bayesian priors used in this work (that is, for constraints from the DESI data alone) are shown in \tabref{priors}.
In summary, we sample the cosmological parameters $H_0$, $\Om$, $\ob$, $\logA$, $\ns$, and $\qbao$, as well as bias parameters in the basis $\left(1+b_1\right) \sigma_8$, $b_2 \sigma_8^2$, $b_s \sigma_8^2$ (chosen to reduce projection effects, described below), \cwrminor{while enforcing a maximization procedure on the counterterms $\alpha_{\left[0,2,4\right]}$ and stochastic parameters $\mathrm{SN}_{\left[0,2,4\right]}$ that efficiently approximates a Jeffreys prior \cite{Jeffreys:1946,Hadzhiyska:2023}.
We note that we sample the parameter $\Om$ instead of $\ocdm$ as in \cite{DESI2024.V.KP5,DESI2024.VII.KP7B}, and we also allow the nuisance parameters $\alpha_{4}$ and $\mathrm{SN}_{4}$ to vary rather than fixing them to zero (the latter corresponds to an older analysis configuration and will cause our parameter errors to be slightly more conservative).
Additionally, use of the Jeffreys prior is unique to this work, and we discuss it in depth at the end of this section.
}
We fix the sum of the neutrino masses, $\sumnu$, to $0.06 \eV.$
\footnote{
\cwrminor{
In \cite{Philcox:2022}, it is noted that $\sumnu$ has some degeneracy with $H_0$, especially in the CMB lensing likelihood. However, they obtained nearly identical constraints when fixing this parameter compared to enforcing their fiducial prior $\sumnu < 0.26 \eV$, i.e. the 2$\sigma$ constraint from the \Planck primary CMB \cite{Planck:2020}.
While we consider their fiducial prior sufficiently conservative, we note that the authors observed degraded constraints when using a prior several times wider, e.g. from the ground-based KATRIN experiment \cite{Aker:2022}.
}
}
We include two external priors: a Big Bang Nucleosynthesis (BBN) prior on $\ob$ \cite{Schoeneberg:2024}, and a wide Gaussian prior on $\ns$ with standard deviation 0.042, corresponding to $10\times$ the width of the posterior reported by the \Planck collaboration using both the primary CMB fluctuations and their gravitational lensing by large-scale structure (TT,TE,EE+lowE+lensing) \cite{Planck:2020}.
When combining our results with CMB lensing, we enforce the narrower $\ns$ prior employed in the \Planck lensing-only analysis: $\ns \sim \mathcal{N}\left(0.96, 0.02^2\right)$ \cite{Planck:2020lens}.
We enforce Gaussian priors on the EFT counterterms $\alpha_{\left[0,2,4\right]}$ such that the $1\sigma$ correction to the power spectrum at $k_{\mathrm{max}}=0.2 \hmpcinv$ is at most 50\% of the total signal; this translates to a $\mathcal{N}\left(0, 12.5^2\right)$ prior \cite{DESI2024.V.KP5}.
\cwrminor{
For the stochastic terms, we enforce a Gaussian prior $\mathcal{N}\left(0, 2^2\right) \times 1/\bar{n}_{\mathrm{g}}$ on the monopole term $\mathrm{SN}_0$, corresponding to a width of $2\times$ the Poissonian shot noise, and $\mathcal{N}\left(0, 5^2\right) \times f_{\mathrm{sat}} \sigma_{1\,\mathrm{eff}}^2/\bar{n}$ for $\mathrm{SN}_2$, corresponding to a width of $5\times$ the characteristic velocity dispersion, where $f_{\mathrm{sat}}$ is the expected satellite galaxy fraction and $\sigma_{1\,\mathrm{eff}}^2$ is the velocity dispersion of these galaxies.
$\mathrm{SN}_4$ uses a prior with the same width as $\mathrm{SN}_2$, but with an additional factor of $\sigma_{1\,\mathrm{eff}}^2$.
}

As mentioned above, we differ from \cite{DESI2024.V.KP5,DESI2024.VII.KP7B} in applying an approximate Jeffreys prior \cite{Jeffreys:1946} to a subset of the parameters comprised by the counterterms and stochastic terms.
Bayesian inference using EFT models, which have a very high-dimensional space of nuisance parameters, is generally prone to parameter biases due to what are known as prior weight effects and prior volume effects, collectively called ``projection effects'' \cite{Simon:2023,Holm:2023}.
The Jeffreys prior, by construction, minimizes sensitivity to the choice of parameterization and therefore also minimizes projection effects.
In \cite{DESI2024.VII.KP7B} it is shown that for full-shape analysis of the DESI 2024 data, projection effects can become pronounced when opening up the parameter space beyond standard \lcdm (which we are doing through the addition of the $\qbao$ parameter).
\cwrminor{
Indeed, when analyzing mock DESI data alone without a Jeffreys prior, we observe $H_0$ offsets at the $\sim$2$\sigma$ level; these offsets are corrected almost entirely after enforcing a Jeffreys prior.
}
While the addition of external datasets can help to mitigate biases due to projection effects \cite{Hadzhiyska:2023,DESI2024.VII.KP7B}, we conservatively choose to apply the Jeffreys prior to further reduce these effects.
We efficiently accomplish this using the technique of \cite{Hadzhiyska:2023}, who showed that the profile likelihood is a good approximation for a nuisance parameter-marginalized posterior in the presence of a Jeffreys prior.
In other words, we simply fix our nuisance parameters (counterterms and stochastic terms) to their best-fit (maximum posterior) values at each point in cosmological+bias parameter space.

\begin{table}
\centering
\begin{tabular}{c c c}
 \hline
 Parameter & Prior & Notes \\
 \hline
 $H_0$ & $\mathrm{Unif}\left[20, 100\right]$ & [$\kmsMpc$] \\
 $\Om$ & $\mathrm{Unif}\left[0.01, 1.0\right]$ & \\ 
 $\ob$ & $\mathcal{N}\left(0.02218, 0.00055^2\right)$ & BBN \cite{Schoeneberg:2024} \\ 
 $\logA$ & $\mathrm{Unif}\left[1.61, 3.91\right]$ & \\ 
 $\ns$ & $\mathcal{N}\left(0.9649, 0.042^2\right)$ & $10\times$ \Planck \cite{Planck:2020} \\
 $\qbao$ & $\mathrm{Unif}\left[0.9, 1.1\right]$ & \\ 
 \hline
 $\left(1+b_1\right) \sigma_8$ & $\mathrm{Unif}\left[0.0, 3.0\right]$ & \\
 $b_2 \sigma_8^2$, $b_s \sigma_8^2$ & $\mathcal{N}\left(0, 5^2\right)$ & \\
 $\alpha_0$, $\alpha_2$, $\alpha_4$ & $\mathcal{N}\left(0, 12.5^2\right)$ & +Approx. Jeffreys prior\tablefootnote{\label{jeffreys_note}Jeffreys prior \cite{Jeffreys:1946} applied in addition to the listed Gaussian prior. In practice, we use an approximation to the Jeffreys prior demonstrated in \cite{Hadzhiyska:2023} (see \secref{priors}).} \\
 $\mathrm{SN}_0$ & $\mathcal{N}\left(0, 2^2\right) \times 1/\bar{n}_{\mathrm{g}}$ & +Approx. Jeffreys prior$^{\ref{jeffreys_note}}$ \\
 $\mathrm{SN}_2$ & $\mathcal{N}\left(0, 5^2\right) \times f_{\mathrm{sat}} \sigma_{1\,\mathrm{eff}}^2/\bar{n}_{\mathrm{g}}$ & +Approx. Jeffreys prior$^{\ref{jeffreys_note}}$ \\ 
 $\mathrm{SN}_4$ & $\mathcal{N}\left(0, 5^2\right) \times f_{\mathrm{sat}} \sigma_{1\,\mathrm{eff}}^4/\bar{n}_{\mathrm{g}}$ & +Approx. Jeffreys prior$^{\ref{jeffreys_note}}$ \\ 
 \hline
\end{tabular}
\caption{
Priors used in this work for constraints from the DESI data alone.
The physically motivated nuisance parameterization used here is described in detail in \cite{DESI2024.V.KP5} and \cite{KP5s2-Maus}.
The prior width of the term $\mathrm{SN}_0$ is multiplied by the Poissonian shot noise $1/\bar{n}_{\mathrm{g}}$, whereas the prior width for the term $\mathrm{SN}_2$ is multiplied by the expected satellite galaxy fraction times the mean satellite velocity dispersion times the Poissonian shot noise, corresponding to the characteristic velocity dispersion \cite{KP5s2-Maus}, and so on for $\mathrm{SN}_4$.
}
\label{table:priors}
\end{table}

\subsection{Emulator}
\label{sec:emulator}

A single evaluation of the full EFT theory can take several seconds, which can be prohibitively time-consuming for a single Markov chain Monte Carlo run (MCMC; see \cite{Robert:2004} and references therein), let alone the many runs needed for testing.
We therefore train an emulator to quickly produce approximate theoretical power spectrum multipoles at any given point in cosmological parameter space.
We train a 4th-order Taylor series emulator for each tracer, a functionality that is built into the \texttt{desilike} code.\footnote{https://desilike.readthedocs.io/en/latest/api/emulators.html\#module-desilike.emulators.taylor}
\cwrminor{
\texttt{desilike} separates the cosmology- and nuisance-dependent parts of the perturbation theory integrals (with the former consuming the bulk of the computation time), so that we only need to Taylor expand in the cosmological parameters.
}

\subsection{Sampling}
\label{sec:sampling}

We run MCMC inference using the \texttt{desilike} implementation of the No-U-Turn sampler (NUTS; \cite{Hoffman:2014}) on the Perlmutter computing cluster at the National Energy Research Scientific Computing Center (NERSC).
We continue sampling until the maximum-eigenvalue Gelman-Rubin statistic \cite{Brooks:1998} satisfies $R-1 < 0.03$.
Because the coverged MCMC chains were generated using the emulated EFT theory, after sampling we chose to recover the true posterior distribution via importance sampling.
That is, we weight each point in the chain by the ratio of the true posterior probability to emulated posterior probability.\footnote{We note that when importance sampling with the true theory, we use the same maximization technique described in \secref{priors} for the counterterms and stochastic terms. That is, at a given point in cosmological+bias parameter space, we use the best-fit counterterms and stochastic terms according to the true posterior.}
We note that after unblinding our analysis, in order to obtain very well-sampled chains for the final combination with several external datasets, we resumed sampling until obtaining an effective sample size (ESS) of \cwrminor{40,000}, defined as the total number of MCMC iterations divided by the integrated correlation length of the chains.
These are the chains that were used to derive the results presented in this work.

\subsection{Validation on Mocks and Unblinding}
\label{sec:validation}

\begin{figure}
\begin{center}
    \includegraphics[width=\columnwidth]{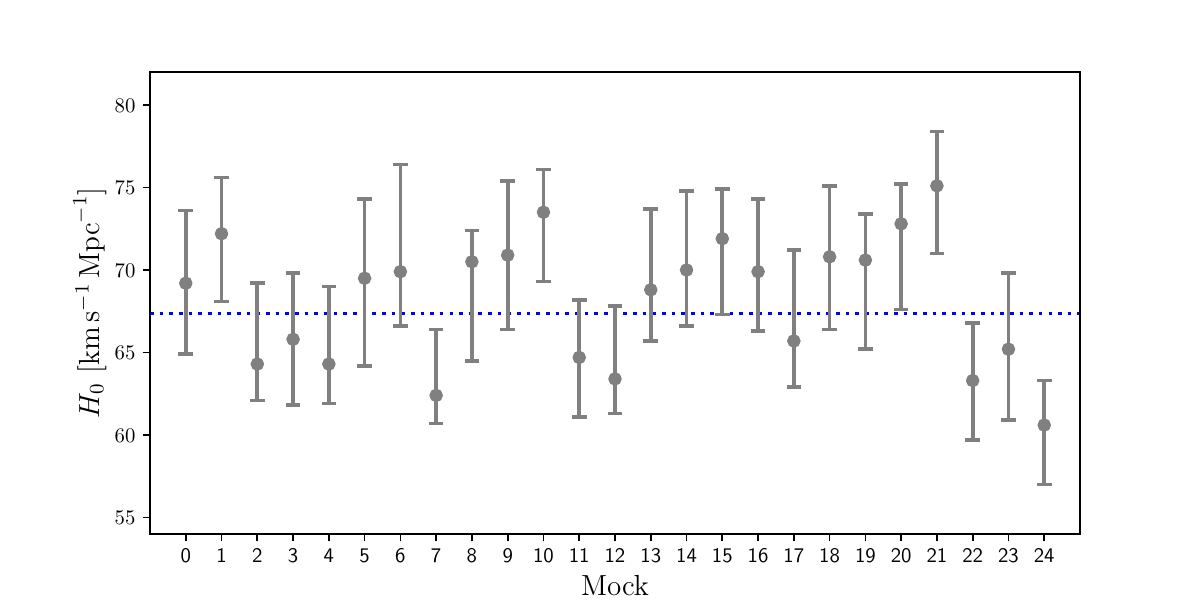}
    \caption{68\% credible-interval constraints on the parameter $H_0$, for all 25 AbacusSummit mocks used for the testing of our analysis pipeline.
    The blue dotted line shows the true underlying $H_0$ value of the mocks.
    18 of the 25 credible intervals (72\%) capture the true $H_0$ value.}
    \label{fig:mocks_h_1d}
\end{center}
\end{figure}

We chose to perform our analysis ``blind'' in order to avoid introducing human bias into the results.
Rather than blinding via a random transformation of the real data (as was done in \cite{DESI2024.V.KP5}), by ``blind'' we mean that we only tested our analysis settings and pipeline using mocks, before finally deciding to run and view our results using the real data.
For our mock testing, we used 25 mock data vectors obtained from AbacusSummit \cite{Maksimova:2021,Garrison:2019} $N$-body mocks, which are based on halo occupation distribution (HOD) models calibrated to match the clustering of the DESI early data release \cite{DESI2023b.KP1.EDR}.
The specific mocks we used for testing are described in \cite{Yuan:2023,Rocher:2024,Smith:2024}.
We used the same covariance matrix described in \secref{covariance}.

Before running our analysis on real data, we ensured that the following criteria were satisfied.
First, we confirmed that among the 25 mock measurements, roughly 68\% of our $1\sigma$ $H_0$ constraints captured the fiducial value used in the mocks.
The results of this test using mock DESI data alone are shown in \figref{mocks_h_1d}, which shows that 18 of the 25 constraints, or 72\%, capture the true $H_0$ value used in the mocks.
Additionally, we observed no significant biases in the parameter constraints when considering two different mock fiber assignment strategies (Fast Fiber Assign \cite{KP3s11-Sikandar} and altMTL \cite{KP3s7-Lasker}).
As a final check before unblinding our results, we ran our analysis pipeline on the real DESI data, and importance sampled the emulated chains using the true theory as described in \secref{sampling}.
We confirmed (only looking at differences in parameter values) that the mean $H_0$ values obtained from the emulated and importance-sampled chains differed by less than $0.3\sigma$.
After satisfying this final criterion, we unblinded our results and combined with external datasets as described in \secref{importance}.
We note that our fiducial datasets for this work, i.e. those chosen before unblinding, were: 1) DESI galaxy clustering; 2) \planckact CMB lensing; and 3) $\Om$ information from each of the DESY5/Pantheon+/Union3 SN-Ia datasets.
After unblinding, we decided to further combine our result with $\Om$ information from the DESI \lya forest as described in \secref{theory}.
Both sets of results are presented in \secref{results}.
\cwr{Additionally, we note that our initial run on real data did not include the clustering of the BGS sample; however, this negligibly affected our main results.}

\subsection{Combination with External Datasets}
\label{sec:importance}

In this paper we combine the base DESI clustering-only analysis with external data at the likelihood level by reweighting our MCMC chains. In the case of CMB lensing, we perform importance sampling using the \planckact likelihood described in \secref{data}. We note that the combination with CMB lensing at the likelihood level is permissible because the covariance between the projected statistics and the multipoles is negligible~\cite{Taylor:2022rgy}. Meanwhile for DESY5 and Pantheon+, we note that the constraints on $\Om$ are Gaussian, leading to a simple analytic re-weighting. However, we find that the Union3 and \lya constraint on $\Om$ is not Gaussian, so in these cases we use {\tt CombineHarvesterFlow}\footnote{https://github.com/pltaylor16/CombineHarvesterFlow}~\cite{Taylor:2024eqc} to find the joint constraints.

\section{Results}
\label{sec:results}

\begin{table}
\centering
\begin{tabular}{c c}
 \hline
 Dataset(s) & $H_0$ [$\kmsMpc$] \\
 \hline
 DESI galaxy clustering only & \cwr{$71.2 \pm 4.1$} \\
 DESI + CMB lensing & \cwr{$70.1^{+2.7}_{-3.3}$} \\
 DESI + CMB lensing + $\Om^{\mathrm{Ly}\alpha\mathrm{AP}}$ & \cwr{$69.6^{+2.4}_{-2.6}$} \\
  DESI + CMB lensing + $\Om^\mathrm{DESY5}$ ~~~~(+ $\Om^{\mathrm{Ly}\alpha\mathrm{AP}}$) & \cwr{$66.1^{+1.8}_{-2.0}$ $\left(66.7^{+1.7}_{-1.9}\right)$} \\
 DESI + CMB lensing + $\Om^\mathrm{Pantheon+}$ (+ $\Om^{\mathrm{Ly}\alpha\mathrm{AP}}$) & \cwr{$67.6^{+1.9}_{-2.2}$ $\left(67.9^{+1.9}_{-2.1}\right)$} \\
 DESI + CMB lensing + $\Om^\mathrm{Union3}$ ~~~~(+ $\Om^{\mathrm{Ly}\alpha\mathrm{AP}}$) & \cwr{$67.3^{+2.1}_{-2.4}$ $\left(67.8^{+2.0}_{-2.2}\right)$} \\
 \hline
\end{tabular}
\caption{$H_0$ constraints obtained in this work. Limits shown are 68\% credible intervals.}
\label{table:results}
\end{table}

\begin{figure}
\begin{center}
    \includegraphics[width=0.75\columnwidth]{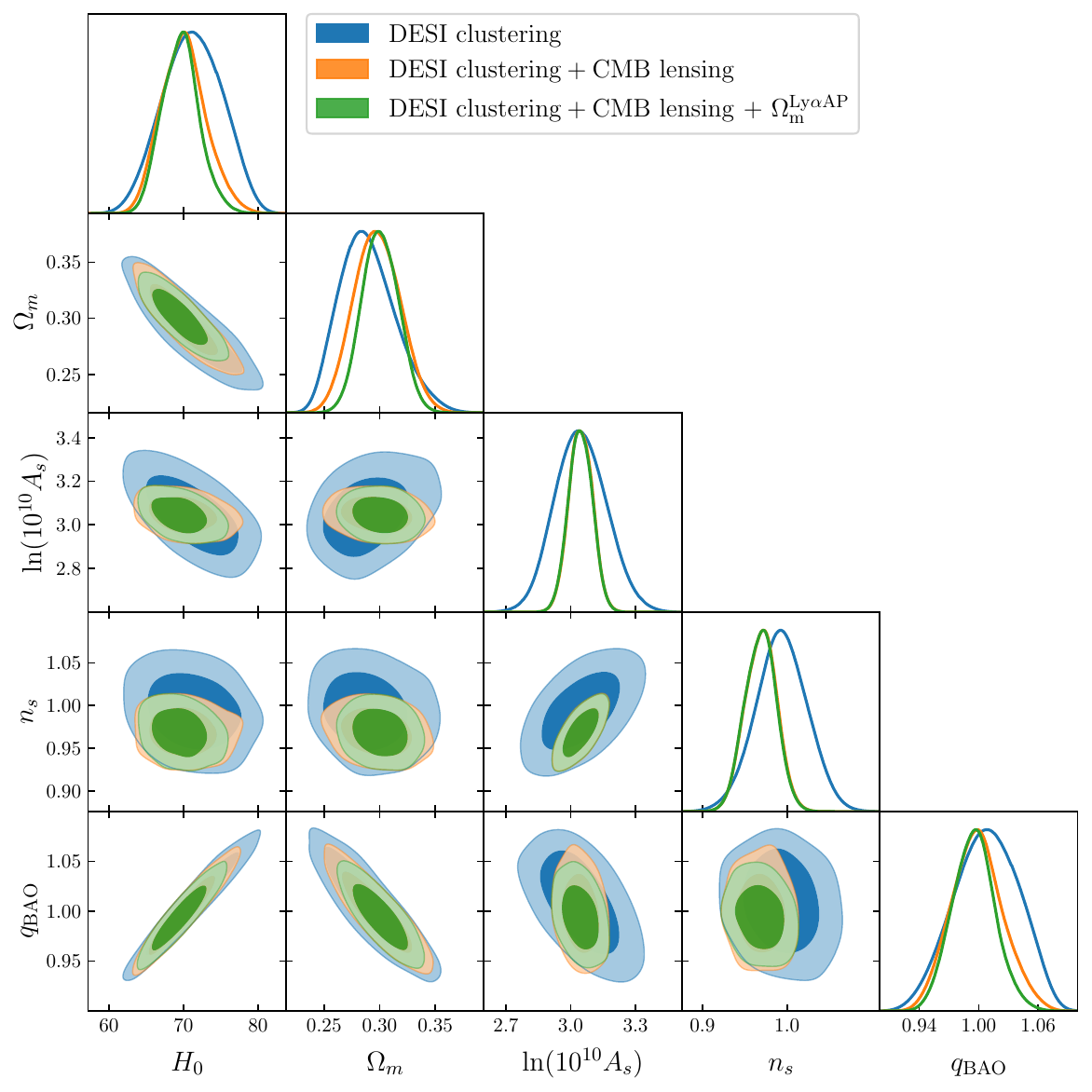}
    \caption{Cosmological parameter constraints from: 1) DESI galaxy clustering only (\cwr{BGS} + LRG + ELG + QSO); 2) DESI clustering + CMB lensing; and 3) DESI clustering + CMB lensing + $\Om^{\mathrm{Ly}\alpha\mathrm{AP}}$.
    CMB lensing refers to the \planckact lensing-only likelihood described in \secref{data}.
    We note that \lya was added after unblinding, as discussed in \secref{validation}.}
    \label{fig:desi_cmblens}
\end{center}
\end{figure}

The results of this work are summarized in \tabref{results}.
In \figref{desi_cmblens} we present our constraints on the parameters $\left\{H_0,~\Om,~\logA,~\ns,~\qbao\right\}$ using the DESI galaxy clustering data alone (\cwr{BGS} + LRGs + ELGs + QSOs), as well as when combining with \planckact CMB lensing and $\Om$ information from the DESI \lya forest ($\Om^{\mathrm{Ly}\alpha\mathrm{AP}}$).
We first note the high quality of the constraints afforded by the DESI data alone, especially considering that we have only analyzed data from the first DESI data release, which corresponds to the first of five total years of observations.
In particular, the DESI clustering-only sound horizon-marginalized $H_0$ constraint, $\cwr{71.2 \pm 4.1} \kmsMpc$, \cwrminor{is already more precise than the analogous constraints obtained from BOSS DR12 in \cite{Farren:2022} and \cite{Philcox:2022}.}
As discussed in \secref{theory}, the \cwr{six} DESI tracers, while individually carrying $\keq$-characteristic degeneracies in the $\Om - H_0$ plane, each intersect to create a more precise joint constraint.
The addition of CMB lensing primarily helps to improve the $H_0$ constraint by intersecting the DESI contour at a slightly steeper angle in the $\Om - H_0$ plane.
To a lesser extent, CMB lensing additionally improves the $H_0$ measurement by better constraining the amplitude $A_{\mathrm{s}}$ of the primordial power spectrum, which also displays some degeneracy with the Hubble constant.
With the addition of $\Om$ information from the DESI \lya forest, we obtain our tightest constraints independent of supernovae (DESI clustering + CMB lensing + $\Om^{\mathrm{Ly}\alpha\mathrm{AP}}$): $H_0 = \cwr{69.6^{+2.4}_{-2.6}} \kmsMpc$.
We finally note that even when allowing the sound horizon rescaling parameter $\qbao$ to vary freely with an uninformative prior, our data broadly seem to prefer values near the \lcdm prediction of $\qbao=1.0$.
This is also the case when combining with CMB lensing and $\Om^{\mathrm{Ly}\alpha\mathrm{AP}}$; the full combination of these datasets gives the constraint $\qbao = \cwr{0.995^{+0.021}_{-0.022}}$.

\begin{figure}
    \centering
    \begin{minipage}[b]{0.49\textwidth}
        \includegraphics[width=\columnwidth]{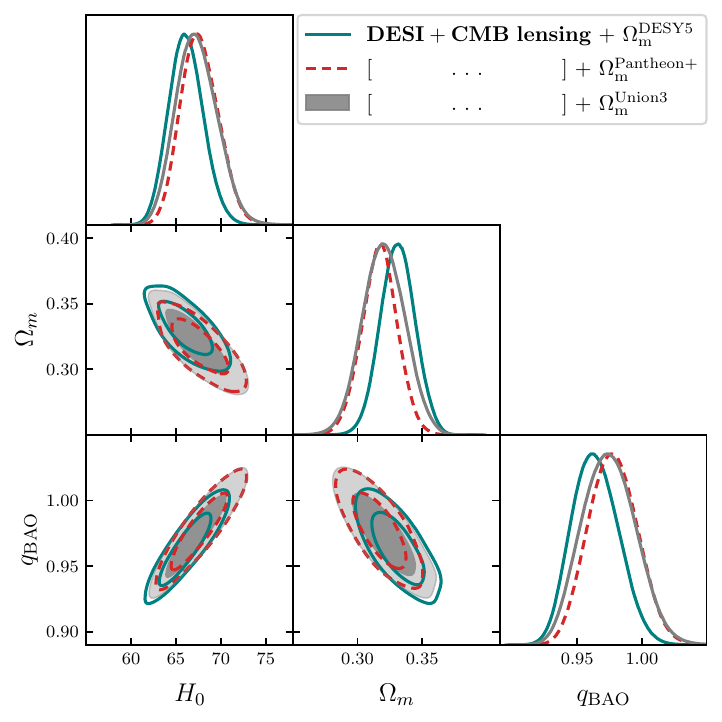}
    \end{minipage}
    \hfill
    \begin{minipage}[b]{0.49\textwidth}
        \includegraphics[width=\columnwidth]{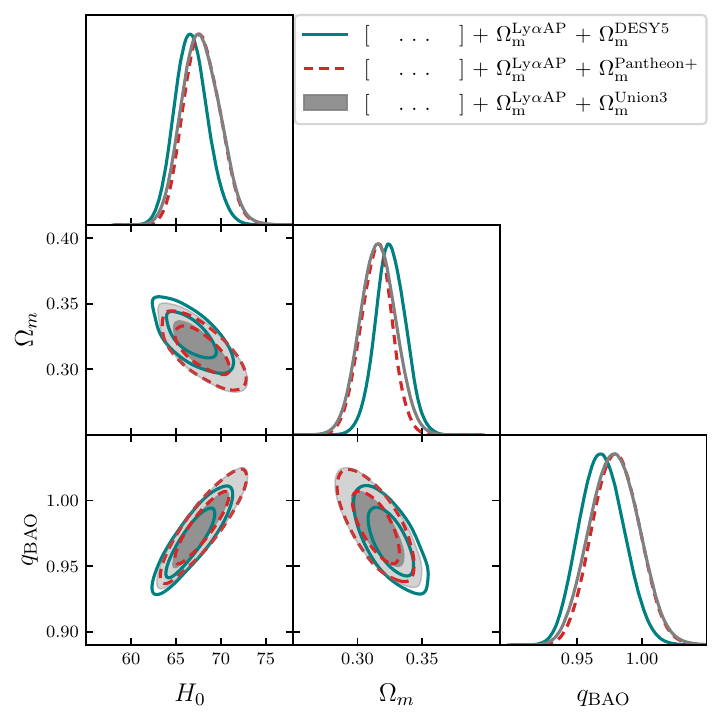}
    \end{minipage}
    \caption{Constraints obtained in this work on the parameters $H_0$, $\Om$, and $\qbao$.
    \textbf{Left:} Constraints from DESI galaxy clustering + CMB lensing + $\Om$ information from each of the DESY5, Pantheon+, and Union3 supernova datasets.
    \textbf{Right:} Same as left, with the addition of $\Om$ information from the DESI \lya forest, as desribed in \secref{theory}.
    We note that \lya was added after unblinding.
    CMB lensing refers to the \planckact lensing-only likelihood described in \secref{data}.}
    \label{fig:main_contours}
\end{figure}

In \figref{main_contours} we show our full constraints on $H_0$, $\Om$, and $\qbao$ using all datasets listed in \secref{data}.
On the left we show the three dataset combinations that were chosen before unblinding: DESI galaxy clustering + CMB lensing + $\Om^{\mathrm{DESY5}}$/$\Om^{\mathrm{Pantheon+}}$/$\Om^{\mathrm{Union3}}$.
On the right we show our tightest constraints, in which we  additionally include $\Om$ information from DESI \lya, again noting that this was decided after unblinding (as discussed in \secref{validation}).
Our tightest $H_0$ constraints obtained in this work are:
\begin{align*}
    H_0 = 
    \begin{cases}
    \begin{aligned}
        ~\cwr{66.7^{+1.7}_{-1.9}}\, \kmsMpc & ~~(\mathrm{DESI~+~CMB~lensing~+~}\Om^{\mathrm{Ly}\alpha\mathrm{AP}}~+~\Om^{\mathrm{DESY5}}) \\[1ex]
        ~\cwr{67.9^{+1.9}_{-2.1}}\, \kmsMpc & ~~(\mathrm{DESI~+~CMB~lensing~+~}\Om^{\mathrm{Ly}\alpha\mathrm{AP}}~+~\Om^{\mathrm{Pantheon+}}) \\[1ex]
        ~\cwr{67.8^{+2.0}_{-2.2}}\, \kmsMpc & ~~(\mathrm{DESI~+~CMB~lensing~+~}\Om^{\mathrm{Ly}\alpha\mathrm{AP}}~+~\Om^{\mathrm{Union3}}) \\[1ex]
    \end{aligned}
    \end{cases}
\end{align*}
This translates to a $\sim$2.9\% average constraint, which as far as the authors are aware is the most precise sound horizon-free $H_0$ measurement to date using a Full Modeling approach.\footnote{We note that \cite{Brieden:2023} obtained a similarly precise constraint using ShapeFit and data from BOSS DR12; their template-based approach and other differences in their analysis pipeline \cwrminor{(in particular their inclusion of ``uncalibrated, unnormalized'' BAO information, which we plan to investigate in future work)} make a direct comparison difficult.}
We can interpret our combined constraints on the parameter $\qbao$ as a test of whether the full dataset is consistent with \lcdm, in the context of sound horizon-modifying new physics.
Because our marginalized posteriors are asymmetric, we adopt the non-Gaussian tension estimation technique of \cite{Raveri:2021}.
We find that our full joint constraints are consistent with \lcdm ($\qbao=1.0$) at up to \tbd{\cwr{1.8}$\sigma$} when including \lya, or \tbd{\cwr{2.0}$\sigma$} without \lya.
\cwrminor{
We note that we observe a slight preference for $\qbao < 1.0$, \refedit{which is driven by the larger $\Om$ preferred by the SN Ia datasets relative to DESI. Loosely speaking, this preferentially selects one side of the $\Om - H_0$ contour in \figref{desi_cmblens}, requiring a decrease in $\qbao$ (i.e. an increase in $\rd$) to compensate the smaller implied $H_0$}.
Future DESI data releases will help to determine whether or not this is simply a statistical fluctuation; \refedit{ it will be particularly important to investigate such deviations in the context of the recent results on dynamical dark energy in DESI DR1 \cite{DESI2024.VI.KP7A,DESI2024.VII.KP7B}.}
}

\refedit{
Finally, we note that because the measurements in this work depend on the broadband shape of the power spectrum, special attention is warranted to determine our sensitivity to large-scale systematics.
In Appendix \ref{sec:systematics}, we demonstrate the robustness of our constraints to several potential sources of such systematic effects.
}

\section{Discussion}
\label{sec:discussion}

\begin{figure}
	\includegraphics[width=\textwidth]{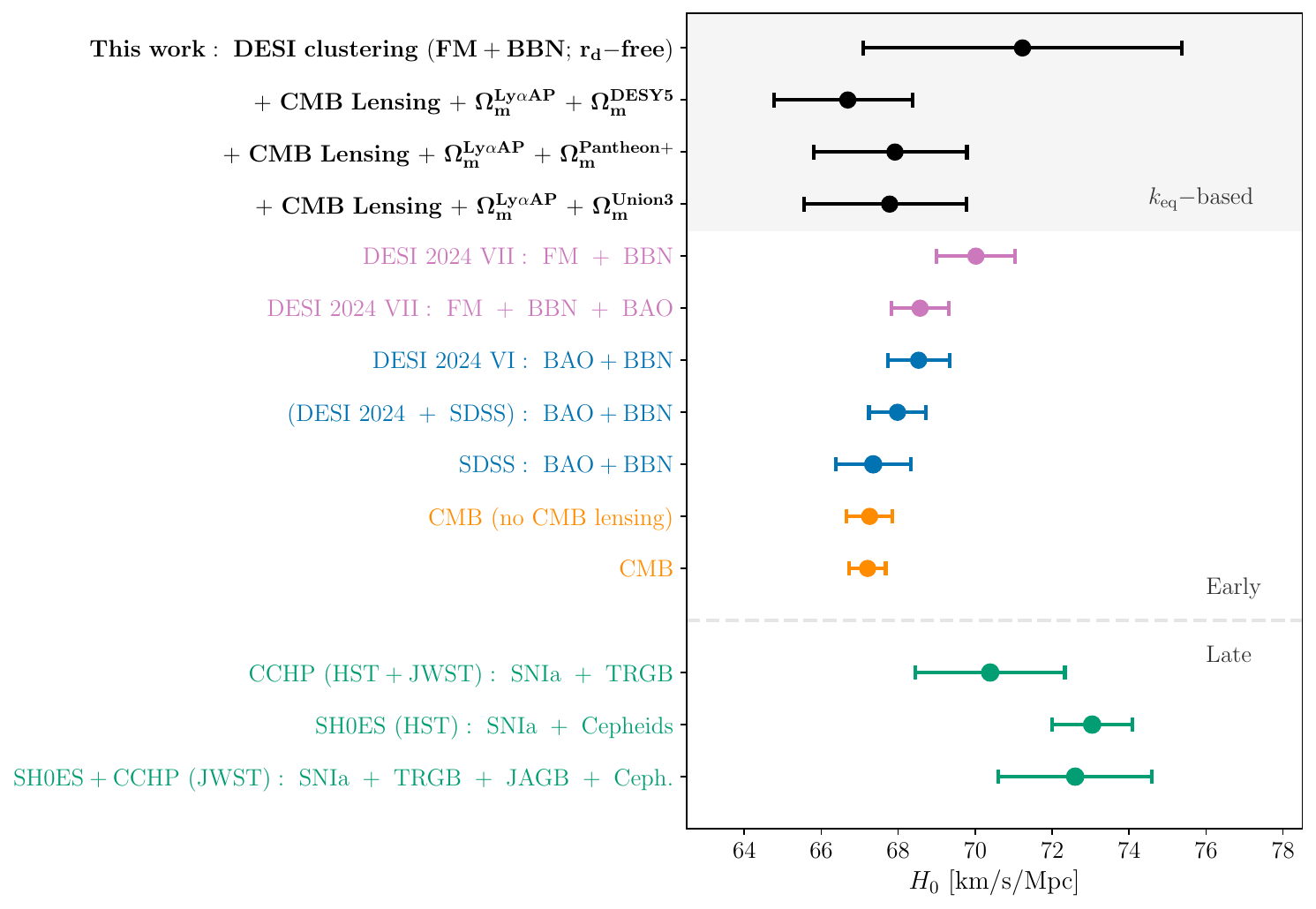}
    \caption{68\% credible interval constraints from various measurements of the Hubble constant.
    The black whiskers with bolded labels show constraints obtained in this work.
    We show our $H_0$ measurement from DESI galaxy clustering alone (\cwr{BGS} + LRG + ELG + QSO), as well as when combined with \planckact CMB lensing and $\Om$ information from the DESI \lya forest and each supernova dataset as discussed in \secref{results}.
    The pink whiskers show constraints from the DESI 2024 full-shape analysis \cite{DESI2024.VII.KP7B}, while the blue whiskers show constraints from the DESI 2024 \cite{DESI2024.VI.KP7A} and/or the Sloan Digital Sky Survey (SDSS; \cite{York:2000}) BAO analyses.
    CMB $H_0$ measurements from the combination of \Planck and ACT data are shown in orange.
    \refedit{
    Finally, the green whiskers at the bottom show local distance ladder $H_0$ constraints from the Chicago-Carnegie Hubble Program (CCHP; \cite{Freedman:2024})\textsuperscript{\textdagger} using data from the Hubble Space Telescope (HST) and the James Webb Space Telescope (JWST), the SH0ES \cite{Riess:2022} team using data from HST, as well as combined constraints from the largest combination of both subsamples available in JWST \cite{Riess:2024}.
    JAGB stands for the J-region Asymptotic Giant Branch method, and TRGB stands for the Tip of the Red Giant Branch calibration method (see the relevant references for further information).
    }}
    \footnotesize \refedit{\textsuperscript{\textdagger} At the time of writing, in \cite{Freedman:2024} the CCHP team quotes their HST+JWST TRGB-calibrated $H_0$ measurement as their main result.}
    \label{fig:h0_whisker}
\end{figure}

In \figref{h0_whisker} we show how the measurements from this work fit into the current landscape of $H_0$ measurements.
We note two interesting observations.
First is that even without using information from the sound horizon scale, our tightest $H_0$ measurements are still in \tbd{\cwr{2.9}}$\sigma$, \tbd{\cwr{2.3}}$\sigma$, or \tbd{\cwr{2.2}}$\sigma$ tension with SH0ES, when combining with DESY5, Pantheon+, and Union3 respectively (again, using the non-Gaussian tension estimation technique of \cite{Raveri:2021}).
Without including \lya (our fiducial data choice made before unblinding), these tensions are \tbd{\cwr{3.0}}$\sigma$, \tbd{\cwr{2.3}}$\sigma$, and \tbd{\cwr{2.3}}$\sigma$.
Second, our measurement is consistent with other early-time measurements that fully or mostly rely on the sound horizon scale.
This would indicate that our measurement is consistent with \lcdm, and in particular we do not find evidence for new early-Universe physics.
We do note, however, that in \secref{results} we found that the dataset combination (DESI clustering + CMB lensing + $\Om^{\mathrm{DESY5}}$) reaches a $\cwr{2.0}\sigma$ deviation from the \lcdm prediction of $\qbao=1.0$, and it will be interesting to see how this changes with the coming three-year and five-year data releases of DESI.

In \cite{Farren:2022}, the authors performed a similar analysis for a simulated Euclid-like survey in an early dark energy (EDE) universe.
Two EDE models were considered; the best-fit EDE model from \Planck + SH0ES data (\cite{Smith:2020}; weaker EDE), and the EDE model favored by ACT (\cite{Hill:2022}; stronger EDE).
They then measured $H_0$ both with and without sound horizon information (assuming a \lcdm fiducial cosmology), finding that EDE generally shifts the sound horizon $H_0$ measurement toward higher values relative to the sound horizon-free measurement, which is expected due to the reduction of the physical sound horizon scale.
The consistency we observe in this work between various early-time $H_0$ measurements, regardless of whether they rely on the sound horizon scale or are independent of it, begins to put pressure on more extreme beyond-\lcdm models such as ACT-favored EDE.\footnote{\refedit{We note that dedicated tests of specific models will be needed to draw rigorous conclusions about their viability.}}
However, we cannot rule out the possibility of less extreme models of new physics (such as \Planck EDE).
The volume probed by galaxy surveys will increase rapidly in the coming years; as more data is collected and the precision of sound horizon-free $H_0$ measurements improves, even stronger constraints will be placed on such models of new early-Universe physics.

\section{Conclusions}
\label{sec:conclusions}

In light of the Hubble tension, there is significant interest in measuring the Hubble constant $H_0$ using independent techniques.
We have performed a precise measurement of $H_0$ without relying on information from the sound horizon scale, using data from the first data release of the DESI survey.
In \secref{theory} we demonstrated our methodology for removing the sound horizon information from our measurement, rescaling and marginalizing over the sound horizon distance using the technique developed in \cite{Farren:2022}.
By jointly analyzing the full-shape clustering of DESI LRGs, ELGs, quasars, \cwr{and the bright galaxy sample} in \cwr{six} redshift bins, we have obtained constraints \cwrminor{that already surpass those from the completed BOSS survey (e.g. \cite{Farren:2022} and \cite{Philcox:2022}).}
When further combining with CMB lensing, uncalibrated type Ia supernovae, and Alcock-Paczy\'{n}ski information from the DESI \lya forest, we have obtained a sub-3\% constraint on $H_0$, the most precise sound horizon-free measurement from LSS to date using a Full Modeling approach.

The complete results from this work were shown in \secref{results}.
Using the full joint dataset including DESI clustering + CMB lensing + $\Om^\mathrm{SN\,Ia}$ + $\Om^{\mathrm{Ly}\alpha\mathrm{AP}}$, our tightest $H_0$ constraints are $\cwr{66.7^{+1.7}_{-1.9}}$, $\cwr{67.9^{+1.9}_{-2.1}}$, and $\cwr{67.8^{+2.0}_{-2.2}} \,\kmsMpc$, when combining with DESY5, Pantheon+, and Union3, respectively.
We highlighted two interesting observations.
The first is that even without including information from the sound horizon scale, our constraints are still in \cwr{2.2-3.0}$\sigma$ tension with the $H_0$ measurement from the SH0ES collaboration \cite{Riess:2022}.
Second, we observed that our measurement is broadly consistent with other early-time $H_0$ measurements that \textit{do} rely on the sound horizon scale.
As discussed in \secref{discussion}, some early-Universe solutions to the Hubble tension (e.g. early dark energy) may lead to deviations in sound horizon-dependent measurements of $H_0$, compared to those that are independent of the sound horizon scale.
Therefore, we have found no significant evidence of new early-Universe physics in this work.
Future data releases from DESI and other ongoing/upcoming galaxy surveys, however, will allow increasingly precise constraints on such models.
It will be exciting to see how this data changes our understanding of the Hubble tension.

\section{Data Availability}

The data used in this analysis will be made public along the Data Release 1 (details in \url{https://data.desi.lbl.gov/doc/releases/}).

\acknowledgments

We thank Oliver Philcox for helpful discussions during this work, as well as Gerrit Farren for thoughtful comments on the manuscript.
We additionally thank Mustapha Ishak-Boushaki and Adam Riess for their comments on the preprint version of this article.

This material is based upon work supported by the U.S.\ Department of Energy (DOE), Office of Science, Office of High-Energy Physics, under Contract No.\ DE–AC02–05CH11231, and by the National Energy Research Scientific Computing Center, a DOE Office of Science User Facility under the same contract. Additional support for DESI was provided by the U.S. National Science Foundation (NSF), Division of Astronomical Sciences under Contract No.\ AST-0950945 to the NSF National Optical-Infrared Astronomy Research Laboratory; the Science and Technology Facilities Council of the United Kingdom; the Gordon and Betty Moore Foundation; the Heising-Simons Foundation; the French Alternative Energies and Atomic Energy Commission (CEA); the National Council of Humanities, Science and Technology of Mexico (CONAHCYT); the Ministry of Science and Innovation of Spain (MICINN), and by the DESI Member Institutions: \url{https://www.desi. lbl.gov/collaborating-institutions}. 

The DESI Legacy Imaging Surveys consist of three individual and complementary projects: the Dark Energy Camera Legacy Survey (DECaLS), the Beijing-Arizona Sky Survey (BASS), and the Mayall z-band Legacy Survey (MzLS). DECaLS, BASS and MzLS together include data obtained, respectively, at the Blanco telescope, Cerro Tololo Inter-American Observatory, NSF NOIRLab; the Bok telescope, Steward Observatory, University of Arizona; and the Mayall telescope, Kitt Peak National Observatory, NOIRLab. NOIRLab is operated by the Association of Universities for Research in Astronomy (AURA) under a cooperative agreement with the National Science Foundation. Pipeline processing and analyses of the data were supported by NOIRLab and the Lawrence Berkeley National Laboratory. Legacy Surveys also uses data products from the Near-Earth Object Wide-field Infrared Survey Explorer (NEOWISE), a project of the Jet Propulsion Laboratory/California Institute of Technology, funded by the National Aeronautics and Space Administration. Legacy Surveys was supported by: the Director, Office of Science, Office of High Energy Physics of the U.S. Department of Energy; the National Energy Research Scientific Computing Center, a DOE Office of Science User Facility; the U.S. National Science Foundation, Division of Astronomical Sciences; the National Astronomical Observatories of China, the Chinese Academy of Sciences and the Chinese National Natural Science Foundation. LBNL is managed by the Regents of the University of California under contract to the U.S. Department of Energy. The complete acknowledgments can be found at \url{https://www.legacysurvey.org/}.

Any opinions, findings, and conclusions or recommendations expressed in this material are those of the author(s) and do not necessarily reflect the views of the U.S.\ National Science Foundation, the U.S.\ Department of Energy, or any of the listed funding agencies.

The authors are honored to be permitted to conduct scientific research on Iolkam Du’ag (Kitt Peak), a mountain with particular significance to the Tohono O’odham Nation.



\bibliographystyle{JHEP}
\bibliography{biblio.bib}


\appendix
\section{Robustness Tests}
\label{sec:systematics}

\refedit{
In this Appendix we demonstrate the robustness of our results to several potential sources of systematic error.
We note that unless otherwise specified, the results shown herein are obtained using emulated theoretical power spectra (see \secref{emulator}), without the additional importance-sampling step described in \secref{sampling}.
}

\subsection{Sky Region/Imaging Systematics Weights}
\label{sec:angular_weights_and_galactic_cap}

\begin{figure}
\begin{center}
    \includegraphics[width=0.5\columnwidth]{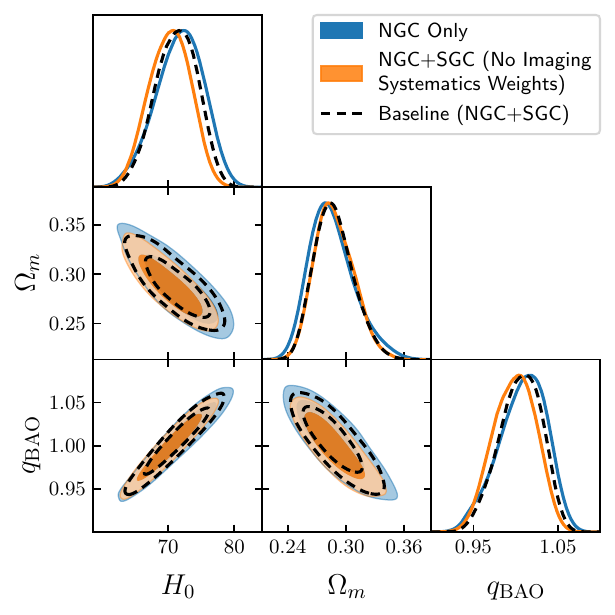}
    \caption{
    \refedit{
    Constraints are shown for analysis variations designed to test our sensitivity to large-scale angular systematics.
    All results displayed are obtained from DESI galaxy clustering alone.
    The baseline results from \secref{results} are shown in the black dashed contours.
    In blue, constraints are shown for only galaxies found in the North Galactic Cap (NGC) sky region.
    In orange are results for the full DESI DR1 footprint (NGC+SGC), but without applying weighting to correct for imaging systematics \cite{DESI2024.II.KP3}.
    In both cases, we find a high degree of consistency with our baseline analysis.
    }}
    \label{fig:angular_weights_and_galactic_cap}
\end{center}
\end{figure}

\refedit{
The constraints obtained in this work are derived from the broadband shape of the galaxy power spectrum, relying more on information from larger scales than in \cite{DESI2024.V.KP5, DESI2024.VII.KP7B}.
This warrants special attention to the treatment of large-scale angular systematics.
We test our robustness to these large-scale angular effects in two ways, with results from DESI galaxy clustering alone shown in \figref{angular_weights_and_galactic_cap}.
One one hand, we check for consistency between different regions of the sky, separately considering the data in the North and South Galactic Caps (NGC/SGC).
As in \cite{DESI2024.V.KP5}, we compare our baseline analysis (NGC+SGC) vs. NGC alone, because the SGC region is too small to be analyzed by itself.
We find very good agreement between the constraints obtained in these two regions, providing confidence in the robustness of our results.
As a second test, we consider the impact of the angular systematics weights used to correct for variations in imaging (see \cite{DESI2024.II.KP3} for a detailed description of this weighting).
We analyze the full data (NGC+SGC) without applying any imaging weights at all, finding that the constraints obtained in this way are not significantly impacted relative to our baseline analysis, indicating that our results are robust to the choice of weighting.
}

\subsection{Minimum \textit{k}-Cutoff}
\label{sec:kmin}

\begin{figure}
\begin{center}
    \includegraphics[width=0.5\columnwidth]{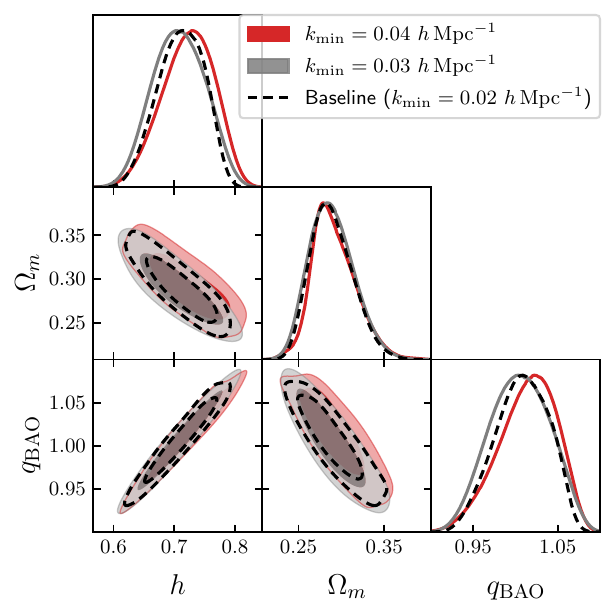}
    \caption{
    \refedit{
    Constraints are shown for various choices of $k_\mathrm{min}$, which sets the large-scale cutoff of our analysis.
    Results are obtained from DESI galaxy clustering alone.
    The baseline results from \secref{results} are shown in the black dashed contours, while the gray and red contours show constraints for $k_\mathrm{min}=0.03\hinvmpc,~0.04\hinvmpc$ respectively.
    We find that our results are robust to the choice of $k_\mathrm{min}$.
    }}
    \label{fig:systest_kmin}
\end{center}
\end{figure}

\refedit{
Motivated by similar reasoning as in the previous section, we test the impact of various choices of $k_\mathrm{min}$, which sets the largest scale probed by our analysis.
Constraints from DESI galaxy clustering alone with $k_\mathrm{min}=0.04 \hinvmpc$ and $0.03 \hinvmpc$, in addition to the baseline choice of $k_\mathrm{min}=0.02 \hinvmpc$, are shown in \figref{systest_kmin}.
We find that the constraints obtained are robust to the choice of $k_\mathrm{min}$, and thus we find no evidence of inconsistencies in the largest-scale measurements.
}

\end{document}